\documentclass[conference]{IEEEtran}
\IEEEoverridecommandlockouts
\usepackage{cite}
\usepackage{graphicx}
\usepackage{amsmath}
\usepackage{times}
\usepackage{latexsym}
\usepackage{bm}
\usepackage{amssymb}
\usepackage[center]{caption2}
\usepackage{array}
\usepackage{fancyhdr}
\usepackage{cite,graphicx,amsmath,amssymb}
\usepackage{citesort}
\usepackage{psfrag}
\usepackage{multirow}

\ifCLASSINFOpdf

\else

\fi
\usepackage[dvipsnames,usenames]{color}
\usepackage[usenames,dvipsnames]{color}

\newtheorem{corollary}{Corollary}
\newtheorem{prop}{Proposition}

\newcommand{\tr}{\text{tr}}

\makeatother

\begin{document}
\title{Secure Massive MIMO Transmission in the Presence
of an Active Eavesdropper}

\author{\IEEEauthorblockN{Yongpeng Wu, Robert Schober, Derrick Wing Kwan Ng, Chengshan Xiao, and Giuseppe Caire}

\thanks{Y. Wu, R. Schober, and D. W. K. Ng are with Institute for Digital Communications, Universit\"{a}t Erlangen-N\"{u}rnberg,
Cauerstrasse 7, D-91058 Erlangen, Germany (Email: yongpeng.wu@lnt.de; schober@lnt.de; kwan@lnt.de). }

\thanks{C. Xiao is with the Department of Electrical and Computer Engineering,
Missouri University of Science and Technology, Rolla, MO 65409, USA (Email: xiaoc@mst.edu). }

\thanks{G. Caire is with Institute for Telecommunication Systems, Technical University Berlin, Einsteinufer 25,
10587 Berlin, Germany (Email: caire@tu-berlin.de). }
}

\maketitle

\begin{abstract}
In this paper, we investigate secure and reliable transmission strategies for multi-cell multi-user massive
multiple-input multiple-output (MIMO) systems in the presence of an active eavesdropper.
We consider a time-division duplex system where uplink training is required and
an active eavesdropper can attack the training phase to cause
pilot contamination  at the transmitter.
This forces the precoder used in the subsequent downlink transmission phase to
implicitly beamform towards the eavesdropper, thus increasing its received signal power.
We derive an asymptotic
achievable secrecy rate for matched filter precoding and artificial noise (AN) generation  at the transmitter
when the number of transmit antennas goes to infinity.
For the achievability scheme at hand, we obtain the optimal power allocation policy
for the transmit signal and the AN in closed form. For the case of correlated fading channels, we show that the impact of the active eavesdropper can be completely removed if the transmit correlation matrices
of the users and the eavesdropper are orthogonal. Inspired by this result, we propose a precoder null space design exploiting
the low rank property of the transmit correlation matrices of massive MIMO channels, which can significantly degrade the eavesdropping
capabilities of the active eavesdropper.
\end{abstract}


\section{Introduction}
Physical layer security  has attracted significant
research interest recently. In  Wyner's pioneering work on information-theoretic security,
a ``wiretap channel'' model was defined and the associated secrecy capacity was obtained \cite{Wyner1975BST}.
More recent studies have investigated the capacity and the precoder design for multiple antenna wiretap channels \cite{Khisti2010TIT_2,Oggier2011TIT,Wu2012TVT}.
When the channel state information (CSI) of the eavesdropper
is imperfectly known at the transmitter, artificial noise (AN) can be generated and transmitted along with the information-bearing
signal to interfere the decoding process of the eavesdropper \cite{Goel2008TWC}.

Most studies on physical layer security assume that perfect CSI of the legitimate channel
is available at the transmitter and do not consider the channel training phase in acquiring the CSI.
However, in time-division duplex (TDD) communication systems,
the base station (BS) needs to estimate the channel for
the subsequent downlink transmission based on the pilot sequences sent by the users in the uplink
training phase.  As a result, a smart eavesdropper might
actively attack this channel training phase by sending the same pilot sequences as the users
to cause pilot contamination at the transmitter.
In this case, the eavesdropping capability can be improved significantly \cite{Zhou2012TWC}.

The emergence of smart mobile devices such as smart phones and  wireless
modems has led to an exponentially increasing demand for wireless data services.
A recent and promising solution to  meet this demand is  massive multiple-input multiple-output (MIMO) technology,
which utilizes {{a very large number of antennas}  at the BS with simple signal processing to serve a comparatively
small (compared to the number of antennas) number of users. The field of massive MIMO communication
systems was initiated by the pioneering work in \cite{Marzetta2008TWC}
which considered multi-cell multi-user TDD communication.
The key idea in \cite{Marzetta2008TWC} is that as the number of transmit antennas
increases,  due to the law of large numbers, the effects of uncorrelated receiver noise
and fast fading vanish.  Then, the only residual interference is
caused by the reuse of the pilot sequences in other cells, which is known as pilot contamination.
Since the publication of \cite{Marzetta2008TWC}, many aspects
of massive MIMO systems have been investigated \cite{Jose2011TWC,Yin2013JSAC,Adhikary2013TIT,Wu2014}.

Physical layer security for massive MIMO systems with
passive eavesdroppers and imperfect CSI has been recently considered in \cite{Chen2014_2}.
Secure massive MIMO transmission for  multi-cell multi-user systems
has been investigated in \cite{Zhu2014},
where a passive eavesdropper attempts to decode the information sent to one of the users.
It was assumed in \cite{Chen2014_2,Zhu2014} that the channel gains of both the desired receiver and the eavesdropper are
 independent and identically distributed (i.i.d.).
The detection of the existence of an active eavesdropper in massive MIMO systems
was investigated in \cite{Im2013} for i.i.d. fading channels.
However, a systematic approach to
combat the pilot contamination attack from the active eavesdropper and maintain secrecy of  communication for general fading channels
was not provided in \cite{Im2013} and has not been studied in the literature, yet.

In this paper, we study secure transmission for TDD multi-cell multi-user massive MIMO systems in the presence
of an active eavesdropper over general fading channels.
We assume that in the uplink training phase,
the active eavesdropper sends the same pilot sequence as the desired receiver to impair the channel estimation
at the transmitter. Therefore, pilot contamination occurs at the transmitter.
Then, the transmitter uses the estimated channel to compute the precoder for downlink transmission.

This paper makes the following key contributions:

\begin{enumerate}

\item  We derive a closed-form expression for the asymptotic
achievable secrecy rate with matched filter precoding and AN generation
at the transmitter assuming the number of transmit antennas tends to infinity.

\item Based on the derived asymptotic expression, we obtain the optimal power allocation policy
for the information signal and the AN in closed form. Moreover, for the special case
of point-to-point transmission over i.i.d. fading channels \cite{Zhou2012TWC},
we prove that the power allocated to the transmit signal should be less than a threshold in order
to ensure reliable secure transmission.

\item For the case of correlated fading channels, we reveal that the impact of the active eavesdropper will
vanish when the signal space (i.e., the span of the channel correlation matrix
eigenvectors corresponding to non-zero eigenvalues) of the users and the eavesdropper are mutually orthogonal.

\item We exploit the low rank property of the transmit correlation matrices of massive MIMO channels \cite{Yin2013JSAC,Adhikary2013TIT}
to design an efficient precoding scheme that transmits in the null space of the transmit correlation matrix of
the eavesdropper.
\end{enumerate}

\emph{Notation:} Vectors  are denoted by lower-case bold-face letters;
matrices are denoted by upper-case bold-face letters. Superscripts $(\cdot)^{T}$, $(\cdot)^{*}$, and $(\cdot)^{H}$
stand for the matrix transpose, conjugate, and conjugate-transpose operations, respectively. We use  ${\tr}({\bf{A}})$ and ${\bf{A}}^{-1}$
to denote the trace operation and the
inverse of matrix $\bf{A}$, respectively.
$\|\cdot \|$ and $| \cdot |$  denote the Euclidean norm of a matrix/vector and
a scalar, respectively. ${\rm{diag}}\left\{\bf{b}\right\}$ denotes a diagonal matrix.
The main diagonal of ${\rm{diag}}\left\{\bf{b}\right\}$ contains the elements of vector $\bf{b}$.
${\rm vec}\left({\mathbf{A}}\right)$ stacks all columns of matrix $\mathbf{A}$ into a vector.
The $M \times M$ identity matrix is denoted
by ${\bf{I}}_M$, and the all-zero matrix is denoted by $\bf{0}$.
The field of complex numbers is denoted
by $\mathbb{C}$ and $E\left[\cdot\right]$ denotes statistical
expectation.
We use $\mathbf{x} \sim \mathcal{CN} \left( {\mathbf{0}_N,{{\bf{R}}}} \right)$ to denote a circularly symmetric complex Gaussian vector
$\mathbf{x} \in {\mathbb{C}^{N  \times 1}}$  with zero mean and covariance matrix ${\bf{R}}$.
 ${\left[ x \right]^ + }$ stands for $\max \left\{ {0,x} \right\}$,
$\otimes$ denotes the Kronecker product, and $ A \mathop \to \limits^{{N} \to \infty } B$ means that
$A$ almost surely converges to $B$ when $N$ goes to infinity.

\section{System Model} \label{sec:multi}
We consider a multi-cell multi-user system with $L + 1$ cells. Each cell contains  a BS
with $N_t$ antennas and $K$ single-antenna users. Without loss of generality,
let the reference cell to be denoted by
$l = 0$.  An active eavesdropper with a single-antenna\footnote{In order to obtain
some initial insights regarding massive MIMO systems with active eavesdropping, we assume
a single-antenna eavesdropper in this paper. It is noted that even a single-antenna active eavesdropper is an extremely
harmful threat to  secrecy transmission since it influences the precoder used in downlink transmission \cite{Zhou2012TWC,Im2013}.
Considering the conclusion in \cite[Appendix B]{Wen2013TIT}, we can also
extend our analysis to  multiple-antenna eavesdroppers. However, the extension to multiple-antenna eavesdroppers requires additional pilot sequence design at the eavesdropper
and a modified precoder design at the transmitter side,
which will be investigated in the journal version of this paper.} is located in the reference cell,
and seeks to recover the private message destined to a specific target user $m$.

\subsection{Uplink Training and Channel Estimation}
In the uplink training and channel estimation phase, the received signal $\mathbf{y}_0$ at the BS
in the reference cell is given by \cite{Yin2013JSAC}
 \begin{small} \begin{align} \label{eq:y_0_multi}
\hspace{-0.3cm} {{\bf{y}}_0} \! = \!  & \sum\limits_{k = 1}^K  \! {\sqrt {{P_{0k}}} \left( {{{\boldsymbol{\omega }}_{0k}} \otimes {{\bf{I}}_{{N_t}}}} \right){\bf{h}}_{0k}^0} \!  + \!  \sum\limits_{l = 1}^L  \! {\sum\limits_{k = 1}^K  \! {\sqrt {{P_{lk}}} \left( {{{\boldsymbol{\omega }}_{lk}} \otimes {{\bf{I}}_{{N_t}}}} \right){\bf{h}}_{lk}^0} } \nonumber \\
& + \sqrt {{P_E}} \left( {{{\boldsymbol{\omega }}_{0m}} \otimes {{\bf{I}}_{{N_t}}}} \right){\bf{h}}_E^0 + {\bf{n}} .
\end{align} \end{small}Here,
$P_{lk}$ and ${{\boldsymbol{\omega}}_{lk}} \in \mathbb{C} {^{\tau  \times 1}}$ are the average transmit power and the pilot sequence of the $k$th user
in the $l$th cell, where $\tau$ denotes the length of the pilot sequence. ${\bf{h}}_{lk}^p = \left( {\bf{R}}_{lk}^p  \right)^{1/2} \mathbf{g}_{lk}^p  \in \mathbb{C}{^{{N_t} \times 1}}$ denotes
the channel between the $k$th user in the $l$th cell and the BS in the $p$th cell,
where $\mathbf{g}_{lk}^p \sim \mathcal{CN} \left( {\mathbf{0}_{N_t},{{\bf{I}}_{{N_t}}}} \right)$ and ${\bf{R}}_{lk}^p  \in {\mathbb{C}^{N_t  \times N_t }}$ is the correlation matrix
of channel ${\bf{h}}_{lk}^p$.  $P_E$ denotes the average transmit power of the eavesdropper.
${\bf{h}}_E^l = \left( {\bf{R}}_{E}^l  \right)^{1/2} \mathbf{g}_{E}^l$ denotes the channel between
the eavesdropper and the BS in the $l$th cell,
where $\mathbf{g}_{E}^l \sim \mathcal{CN} \left( {\mathbf{0}_{N_t},{{\bf{I}}_{{N_t}}}} \right)$ and ${\bf{R}}_{E}^l
\in {\mathbb{C}^{N_t  \times N_t }} $ is the correlation matrix of channel ${\bf{h}}_E^l$.
$ \mathbf{n} \in {\mathbb{C}^{\tau N_t  \times 1}}$  is a zero-mean complex Gaussian noise vector
with covariance matrix $N_0 \mathbf{I}_{\tau N_t}$. We assume that the same $K$ orthogonal pilot sequences
are used for the $K$ users in each cell \cite{Marzetta2008TWC}, i.e.,
${{\boldsymbol{\omega }}_{0k}} = {{\boldsymbol{\omega }}_{1k}} =  \cdots {{\boldsymbol{\omega }}_{Lk}} = {{\boldsymbol{\omega }}_k}$, ${\boldsymbol{\omega }}_{lk}^H{{\boldsymbol{\omega }}_{lk}} = \tau ,
{\boldsymbol{\omega }}_{lk}^H{{\boldsymbol{\omega }}_{lp}} = 0$. Then,
 the minimum mean square error (MMSE) estimate of ${\bf{h}}_{0m}^0$  is given by \cite{Kailath2000}
 \begin{small} \begin{multline} \label{eq:mmse_h0m}
{\widehat {\bf{h}}_{0m}^0} = \\
\sqrt {{P_{0m}}} {\bf{R}}_{0m}^0{\left( \! {{N_0}{{\bf{I}}_{{N_t}}} \! + \! \tau \left( {\sum\limits_{t = 0}^L {{P_{tm}}{\bf{R}}_{tm}^0} \! + \! {P_E}{{\bf{R}}_E^0}} \right)} \! \right)^{ - 1}}{\widetilde {\bf{y}}_{0m}},
\end{multline} \end{small}where
 \begin{small} \begin{align} \label{eq:y_0m}
{\widetilde {\bf{y}}_{0m}} \!  = \! \sqrt {{P_{0m}}} \tau {\bf{h}}_{0m}^0 \! + \! \sum\limits_{t = 1}^L {\sqrt {{P_{tm}}} \tau {\bf{h}}_{tm}^0}  \! + \! \sqrt {{P_E}} \tau {\bf{h}}_E^0 \! + \!{\left( {{{\boldsymbol{\omega }}_m} \! \otimes \! {{\bf{I}}_{{N_t}}}} \right)^H}{\bf{n}}.
\end{align} \end{small}Here, ${\bf{h}}_{0m}^0 = {\bf{\widehat h}}^0_{0m} + {\bf{e}}_{0m}^0$, where the estimated channel ${\bf{\widehat h}}^0_{0m} \sim \mathcal{CN} \left({\mathbf{0}_{N_t}, {\widehat {\bf{R}}_{0m}^0} }\right)$
and the estimation error ${\bf{e}}_{0m}^0 \sim \mathcal{CN} \left({\mathbf{0}_{N_t}, {{\bf{R}}_{0m}^0}  -  {\widehat {\bf{R}}_{0m}^0}}\right)$ are mutually independent. The correlation
matrix ${\widehat {\bf{R}}_{0m}^0}$ is given by
 \begin{small} \begin{multline} \label{eq:R0m0_est}
\widehat {\bf{R}}_{0m}^0 = \\
{P_{0m}}\tau {\bf{R}}_{0m}^0{\left( {{N_0}{{\bf{I}}_{{N_t}}} \! + \! \tau \left( {\sum\limits_{t = 0}^L {{P_{tm}}{\bf{R}}_{tm}^0} \! + \! {P_E}{\bf{R}}_E^0} \right)} \right)^{ - 1}}{\bf{R}}_{0m}^0.
\end{multline} \end{small}

Similarly, the MMSE channel estimates for the $m$th user and the $k$th user, $k = 1,2,\cdots,K, k\neq m$, in the $l$th cell, $l = 0,1,\cdots,L$, are given by
 \begin{small} \begin{multline} \label{eq:hnmn_est}
\widehat {\bf{h}}_{lm}^l = \\
 \sqrt {{P_{lm}}} {\bf{R}}_{lm}^l{\left( {{N_0}{{\bf{I}}_{{N_t}}} \! + \! \tau \left( {\sum\limits_{t = 0}^L {{P_{tk}}{\bf{R}}_{tm}^l} \! + \! {P_E}{\bf{R}}_E^l} \right)} \right)^{ - 1}}{\widetilde {\bf{y}}_{lm}}
\end{multline} \end{small}
 \begin{small} \begin{align} \label{eq:y_nn}
{\widetilde {\bf{y}}_{lm}} = & \sqrt {{P_{lm}}} \tau {\bf{h}}_{lm}^l + \sum\limits_{t = 0, t \neq l }^L {\sqrt {{P_{tm}}} \tau {\bf{h}}_{tm}^l} \nonumber \\
& + \sqrt {{P_E}} \tau {\bf{h}}_E^l + {\left( {{{\boldsymbol{\omega}}_m} \otimes {{\bf{I}}_{{N_t}}}} \right)^H}{\bf{n}}
\end{align} \end{small}and
\begin{small} \begin{align} \label{eq:hnnk_est}
{\widehat {\bf{h}}_{lk}}^l & = \sqrt {{P_{lk}}} {\bf{R}}_{lk}^l{\left( {{N_0}{{\bf{I}}_{{N_t}}} + \tau \sum\limits_{t = 0}^L {{P_{tk}}{\bf{R}}_{tk}^l} } \right)^{ - 1}}{\widetilde {\bf{y}}_{lk}} \\
{\widetilde {\bf{y}}_{lk}} & = \sqrt {{P_{lk}}} \tau {\bf{h}}_{lk}^l + \sum\limits_{t = 0,t \ne l}^L {\sqrt {{P_{tm}}} \tau {\bf{h}}_{tk}^l}  + {\left( {{{\boldsymbol{\omega }}_k} \otimes {{\bf{I}}_{{N_t}}}} \right)^H}{\bf{n}},
\end{align} \end{small}respectively.
The correlation matrices of $\widehat {\bf{h}}_{lm}^l $ and ${\widehat {\bf{h}}_{lk}}^l$  are given by
 \begin{small} \begin{equation}
\widehat {\bf{R}}_{lm}^l =
{P_{lm}}\tau {\bf{R}}_{lm}^l{\left( {{N_0}{{\bf{I}}_{{N_t}}} + \tau \left( {\sum\limits_{t = 0}^L {{P_{tk}}{\bf{R}}_{tm}^l}  + {P_E}{\bf{R}}_E^l} \right)} \right)^{ - 1}}{\bf{R}}_{lm}^l \label{eq:Rnmn_est}
\end{equation} \end{small}and
 \begin{small} \begin{equation}
\widehat {\bf{R}}_{lk}^l = {P_{lk}}\tau {\bf{R}}_{lk}^l{\left( {{N_0}{{\bf{I}}_{{N_t}}} + \tau \sum\limits_{t = 0}^L {{P_{tk}}{\bf{R}}_{tk}^l}  } \right)^{ - 1}}{\bf{R}}_{lk}^l,
\end{equation} \end{small}respectively.

{\emph{Remark 1:}} We assume that the correlation matrices
of the users and the eavesdropper are perfectly known at the transmitter, see (\ref{eq:mmse_h0m}), (\ref{eq:hnmn_est}), (\ref{eq:hnnk_est}).
For massive MIMO systems, it is reasonable to assume that the
statistical CSI of the users of the system is known at the BS \cite{Yin2013JSAC}.
Therefore, if the BS attempts to transmit a private message to some users and treats other users as eavesdroppers, i.e.,
the eavesdropper is an idle user of the system, the correlation matrix of the eavesdropper can also assumed
to be known.
In other cases, the assumption that the correlation matrices of the active
eavesdropper are available at the transmitter may also be reasonable.
For example, we can
obtain $E\left[{\widetilde {\bf{y}}_{lm}}{\widetilde {\bf{y}}_{lm}}^H\right]$ by averaging
${\widetilde {\bf{y}}_{lm}}$ over different data slots.
Eq. (\ref{eq:y_nn}) suggests that $E\left[{\widetilde {\bf{y}}_{lm}}{\widetilde {\bf{y}}_{lm}}^H\right]$ is
the sum of the correlation matrices of all users and the eavesdropper.
Then, ${P_E}{\bf{R}}_E^l$ can be obtained by
subtracting the correlation matrices of the legitimate users
and the noise from $E\left[{\widetilde {\bf{y}}_{lm}}{\widetilde {\bf{y}}_{lm}}^H\right]$.

\subsection{Downlink Data Transmission}
Next, we consider data transmission. We assume that the BSs in
all $L + 1$ cells perform jamming to prevent eavesdropping in their own cell.
Then, the transmit signal in the $l$th cell $l = 0,1,\cdots,L$ is given by
\begin{small} \begin{align}\label{eq:xn}
{{\bf{x}}_l} = \sqrt P \left( {\sqrt p \sum\limits_{k = 1}^K {{{\bf{w}}_{lk}}{s_{lk}}}  + \sqrt q {{\bf{U}}_{{\rm null},\,l}}{{\bf{z}}_l}} \right),
\end{align} \end{small}where
$P$ is the average transmit power for downlink transmission and
$s_{lk}$ is the transmit signal for the $k$th user in the $l$th cell with
$E\left[|s_{lk}|^2\right] = 1$.  $p$ and $q$ represent the power allocation
between the transmit signal and the AN with $p +  q =1$.
Due to the high implementation complexity of
the matrix inversion required for zero forcing  and MMSE precoding, here we adopt the
simple matched filter precoding for massive MIMO systems \cite{Marzetta2008TWC,Jose2011TWC,Zhu2014}. Thus,
we set ${{\bf{w}}_{lk}} = \frac{{\widehat {\bf{h}}_{lk}^l}}{{\left\| {\widehat {\bf{h}}_{lk}^l} \right\|}}$ for the $k$th user in the $l$th cell.
${\bf{U}}_{{\rm null},\,l}$ and $\mathbf{z}_l \sim \mathcal{CN} \left( {\mathbf{0}_{N_t}, {{\bf{I}}_{{N_t}}}} \right)$
 denote the AN shaping matrix and the AN vector in the $l$th cell, respectively. We define $\widehat {\bf{H}}_l^l
 = \left[\mathbf{\widehat {\bf{h}}}_{l1}^l, \mathbf{\widehat {\bf{h}}}_{l2}^l,\cdots, \mathbf{\widehat {\bf{h}}}_{lK}^l\right]$.
 To reduce the implementation complexity when the number of antennas is large,
we fix the AN shaping matrix to be the asymptotic null space of $\widehat {\bf{H}}_l^l$.
Based on \cite[Corollary 1]{Evans2000TIT}, we have $ \frac{1}{N_t}{\left( {\widehat {\bf{H}}_l^l} \right)^H}\widehat {\bf{H}}_l^l \!\mathop
 \to  \limits^{{N_t} \to \infty } \! \!
\frac{1}{N_t}{\rm diag} \! {\left[{\tr\left( {\widehat {\bf{R}}_{l1}^l} \right)},  \! {\tr\left( {\widehat {\bf{R}}_{l2}^l} \right)},\!\cdots,\! {\tr\left( {\widehat {\bf{R}}_{lK}^l} \right) } \!\right]}$.
Thus, we set ${{\bf{U}}_{{\rm null},\,l} } = {{\bf{I}}_{{N_t}}} - \widehat {\bf{H}}_l^l{\rm diag}\left[{\tr\left( {\widehat {\bf{R}}_{l1}^l} \right)^{-1}}, {\tr\left( {\widehat {\bf{R}}_{l2}^l} \right)^{-1}},\cdots, {\tr\left( {\widehat {\bf{R}}_{lK}^l} \right)^{-1} }\right]{\left( {\widehat {\bf{H}}_l^l} \right)^H}$.
It can be shown that $ \frac{1}{N_t}\tr\left( {{\bf{U}}_{{\rm null},\,l} }  {{\bf{U}}^H_{{\rm null},\,l} }\right) \mathop  \to \limits^{{N_t} \to \infty } \frac{1}{N_t} (N_t - K)$.
To ensure $ \frac{1}{N_t} \mathbf{x}_l^H  \mathbf{x}_l \mathop  \to \limits^{{N_t} \to \infty } \frac{1}{N_t} P$,  we set $K p + (N_t - K)  q = 1$.

The received signal  at the $m$th user in the reference cell, ${y_{0m}}$, and at the eavesdropper, $y_{\rm eve}$,
are given by
{\small
 \begin{align}
{y_{0m}}  = \sum\limits_{l = 0}^L {{{\left( {{\bf{h}}_{0m}^l} \right)}^H}{{\bf{x}}_l}}  + {{n}}_{0m} \label{eq:y_0m}
\end{align} }and
\begin{small}
\begin{align}
{y_{\rm eve}} = \sum\limits_{l = 0}^L {{{\left( {{\bf{h}}_{E}^l} \right)}^H}{{\bf{x}}_l}}  + {{n}}_{\rm eve}, \label{eq:y_eve}
\end{align}
 \end{small}respectively.
Here, ${{n}}_{0m}$ and ${{n}}_{\rm eve}$ are zero-mean Gaussian noise processes with variance $N_{0,\rm{d}}$.
We define the signal-to-noise ratio (SNR) of the downlink transmission as $\gamma = P/ N_{0,\rm{d}}$.

\section{Matched Filter Precoding and AN Generation}
An achievable ergodic secrecy rate of massive MIMO systems in Section II
can be expressed as \cite{Zhu2014}
\vspace{0.1cm}
\begin{small}  \begin{align}\label{eq:R_sec}
{R_{\sec}} = {\left[ {{R_{0m}} - {C_{\rm{eve}}} } \right]^ + }
 \end{align} \end{small}where $R_{0m}$ and $C_{\rm{eve}}$ denote the achievable ergodic rate between the BS and the $m$th user
and  the ergodic capacity between the BS and the eavesdropper,  in the $0$th cell, respectively.  The
achievable ergodic rate ${R_{0m}}$ is given by \cite[Eq. (8)]{Zhu2014}
\vspace{0.1cm}
\begin{small}
 \begin{align}\label{eq:R_B}
{R_{0m}} = E\left[{\log _2}\left( {1 + {\rm SINR}_{0m}} \right) \right].
 \end{align}
 \end{small}
\vspace{0.1cm} \\
 where ${\rm SINR}_{0m}$ is given by
\begin{small} \begin{align}\label{eq:SINR_0m}
{\rm SINR}_{0m} = \frac{p \gamma{{ { { \left| {{{\left( {{\bf{h}}_{0m}^0} \right)}^H}{{\bf{w}}_{0m}}}  \right|}}^2}}}{A}
\end{align} \end{small}
\begin{small} \begin{align}\label{eq:SINR_0m_b}
A & \! = \! p \gamma \sum\limits_{k = 1,k \ne m}^K {\left[ {{{\left| {{{\left( {{\bf{h}}_{0m}^0} \right)}^H}{{\bf{w}}_{0k}}} \right|}^2}} \right]} \! + \! q \gamma \left[ {{{\left| {{{\left( {{\bf{h}}_{0m}^0} \right)}^H}{{\bf{U}}_{{\rm null},\,0}}} \right|}^2}} \right] \! + \! p\gamma  \nonumber \\
& \hspace{-0.3cm} \! \times \! \sum\limits_{l = 1}^L {\left[ {\sum\limits_{k = 1}^K {\left[ {{{\left| {{{\left( {{\bf{h}}_{0m}^l} \right)}^H}{{\bf{w}}_{lk}}} \right|}^2}} \right]} } \!+ \! q \gamma \sum\limits_{l = 1}^L {\left[ {{{\left| {{{\left( {{\bf{h}}_{0m}^l} \right)}^H}{{\bf{U}}_{{\rm null},\,l}}} \right|}^2}} \right]}   \right]} \! + \! 1.
\end{align} \end{small}

To capture the worst case, we assume that
the eavesdropper has perfect knowledge of its own channel and is able to decode and
cancel the signals of all intra-cell and inter-cell users from
the received signal ${\mathbf{y}_{\rm eve}}$ in (\ref{eq:y_eve}) except for the signal intended for the $m$th user in the $0$th cell.
In this case, ${C_{\rm{eve}}}$ can be expressed as \cite[Eq. (7)]{Zhu2014}

\begin{align}\label{eq:R_B}
{C_{\rm{eve}}} = E\left[ {\log _2}\left( {1 + {\rm SINR}_{\rm eve}} \right) \right]
\end{align}
\begin{small} \begin{align}\label{eq:SINR_E_mul}
{\rm SINR}_{\rm eve} = \frac{{p \gamma {{\left( {{\bf{h}}_E^0} \right)}^H}{{\bf{w}}_{0m}}{{ {{{\bf{w}}_{0m}^H}}}}{\bf{h}}_E^0}}{B}
\end{align} \end{small}where
\begin{small} \begin{align}\label{eq:eve_B}
B  \! \!= \!\!   q  \gamma  {{\left(\! {{\bf{h}}_E^0} \!\right)\!^H}}\!{{\bf{U}}_{{\rm null},\,0}}{{{{{\bf{U}}_{{\rm null},\,0}^H }}}}{\bf{h}}_E^0  \! + \! q  \gamma \! \sum\limits_{l = 1}^L \! {{{\left(\! {{\bf{h}}_E^l} \!\right)\!^H}}\!{{\bf{U}}_{{\rm null},\,l}}{{ {{{\bf{U}}_{{\rm null},\,l}^H}} }}{\bf{h}}_E^l} \! + \! 1.
\end{align} \end{small}

Now, we are ready to provide an asymptotic achievable secrecy rate expression
when the number of antennas tends to infinity, which is given in the following proposition.
\begin{prop}\label{prop:sec_rate_mul}
An asymptotic achievable secrecy rate for matched filter precoding and AN generation
for multi-cell multi-user massive MIMO systems with an active eavesdropper
is given by
 \begin{small} \begin{align}\label{eq:asy_sec}
{R_{\sec, \, \rm{asy}}} & \mathop  \rightarrow \limits^{{N_t} \to \infty } \left[  {\log _2}\left( {1 + {\rm SINR}_{0m,\,{\rm asy}}} \right) \right.  \nonumber \\
& \left. - {\log _2}\left( {1 + {\rm SINR}_{{\rm eve,\,asy}}} \right) \right]^{+},
\end{align} \end{small}where
\begin{small} \begin{align}
{\rm SINR}_{0m,\,{\rm asy}} &= \frac{{ p \gamma {\theta _m}}}{{p\gamma {\theta _{b,p}} +  q\gamma {\theta _{b,q}} + 1}}  \label{eq:b} \\
{\rm SINR}_{\rm eve,\, asy} &= \frac{{p\gamma }{\theta _{e,e}}}{{q\gamma {\theta _{e,q}} + 1}}  \label{eq:e}
\end{align} \end{small}with
\begin{small} \begin{align}
\hspace{-0.4cm} {\theta _m} = \tr\left( {\widehat {\bf{R}}_{0m}^0} \right) + \tr\left( {\widehat {\bf{R}}_{0m}^0} \right)^{-1} {{\tr\left({\left({{\bf{R}}_{0m}^0 - \widehat{\bf{R}}_{0m}^0} \right) \widehat{\bf{R}}_{0m}^0} \right)}}
\end{align} \end{small}
 \begin{small} \begin{multline}
\hspace{-0.3cm} {\theta _{b,p}} = \! \sum\limits_{l = 0}^L {\sum\limits_{k = 1,k \ne m}^K {\tr{{\left( {\widehat {\bf{R}}_{lk}^l} \right)}^{ - 1}}\!
\tr\left( \!{{\bf{R}}_{0m}^l\widehat {\bf{R}}_{lk}^l} \!\right)}}
 \! + \! \sum\limits_{l = 1}^L { \!\tr{{\left(\! {\widehat {\bf{R}}_{lm}^l} \!\right)}^{ - 1}} \! \Lambda _{0m}^l}
\end{multline} \end{small}
 \begin{small} \begin{align}
\hspace{-0.6cm} {\theta _{b,q}} & = \sum\limits_{l = 0}^L {\tr\left( {{\bf{R}}_{0m}^l} \right)}  - \sum\limits_{l = 0}^L \sum\limits_{k = 1,k \ne m}^K {\tr{{\left( {\widehat {\bf{R}}_{lk}^l} \right)}^{ - 1}}\tr\left( {{\bf{R}}_{0m}^l\widehat {\bf{R}}_{lk}^l} \right)} \nonumber \\
& - \sum\limits_{l = 0}^L {\tr{{\left( {\widehat {\bf{R}}_{lm}^l} \right)}^{ - 1}}\Lambda _{0m}^l}
\end{align} \end{small}
 \begin{small} \begin{align}
\Lambda _{0m}^l  & = {\tau ^2}{P_{0m}}{\left| {\tr\left( {{\bf{C}}_{lm}^l{\bf{R}}_{0m}^l} \right)} \right|^2}  + \tau {N_0}\tr\left( {{\bf{R}}_{0m}^l{\bf{C}}_{lm}^l{{\left( {{\bf{C}}_{lm}^l} \right)}^H}} \right) \nonumber \\
  & + {\tau ^2}\sum\limits_{t = 1}^L {{P_{tm}}\tr\left( {{\bf{R}}_{0m}^l{\bf{C}}_{lm}^l{\bf{R}}_{tm}^l{{\left( {{\bf{C}}_{lm}^l} \right)}^H}} \right)} \nonumber \\
  & + {\tau ^2}{P_E}\tr\left( {{\bf{R}}_{0m}^l{\bf{C}}_{lm}^l{\bf{R}}_E^l{{\left( {{\bf{C}}_{lm}^l} \right)}^H}} \right)
\end{align} \end{small}
 \begin{small} \begin{equation}
{\bf{C}}_{lm}^l \! = \! \sqrt {{P_{lm}}} {\bf{R}}_{lm}^l{\left( {{N_0}{{\bf{I}}_{{N_t} }} \! + \! \tau \left( {\sum\limits_{t = 0}^L {{P_{tm}}{\bf{R}}_{tm}^l} \!  + \! {P_E}{{\bf{R}}_E^{l}}} \right)} \right)^{ - 1}}.
\end{equation} \end{small}
 \begin{small} \begin{equation}
\hspace{-2.9cm} {\theta _{e,e}}  = \frac{\Lambda _E^0}{\tr\left( {\widehat {\bf{R}}_{0m}^0} \right)} \hspace{2.9cm}
\end{equation} \end{small}
 \begin{small} \begin{align}
\hspace{-0.3cm} {\theta _{e,q}} & =  \sum\limits_{l = 0}^L \tr\left( {{\bf{R}}_E^l} \right) - \sum\limits_{l = 0}^L {\sum\limits_{k = 1,k \ne m}^K {\tr{{\left( {\widehat {\bf{R}}_{lk}^l} \right)}^{ - 1}}\tr\left( {{\bf{R}}_E^l\widehat {\bf{R}}_{lk}^l} \right)} } \nonumber \\
& - \sum\limits_{l = 0}^L {\tr{{\left( {\widehat {\bf{R}}_{lm}^l} \right)}^{ - 1}}\Lambda _E^l}
\end{align} \end{small}
 \begin{small} \begin{align}
\Lambda _E^l & =  {\tau ^2}{P_E}{\left| {\tr\left( {{\bf{C}}_{lm}^l{\bf{R}}_E^l} \right)} \right|^2}  + \tau {N_0}\tr\left( {{\bf{R}}_E^l{\bf{C}}_{lm}^l{{\left( {{\bf{C}}_{lm}^l} \right)}^H}} \right) \nonumber \\
 &+ {\tau ^2} \sum\limits_{t = 0}^L {{P_{tm}}\tr\left( {{\bf{R}}_E^l{\bf{C}}_{lm}^l{\bf{R}}_{tm}^l{{\left( {{\bf{C}}_{lm}^l} \right)}^H}} \right)}.
\end{align} \end{small}
\begin{proof}
See Appendix \ref{proof:prop:sec_rate_mul}.
\end{proof}
\end{prop}

Based on Proposition \ref{prop:sec_rate_mul}, we can obtain several new insights for massive MIMO systems with an active eavesdropper,
which are summarized in the following corollaries. The proofs of these corollaries can be directly obtained by finding feasible $p$ which maximize
${R_{\sec, \, \rm{asy}}}$ or yield ${R_{\sec, \, \rm{asy}}} > 0$ in (\ref{eq:asy_sec}).

\begin{corollary} \label{coro:opt_power}
Let us define
\begin{small} \begin{align}\label{eq:p1}
& {p_1} = \nonumber \\
&\frac{{ \! - \!\left( {{a_1}{c_2}  \! - \!  {a_2}{c_1}} \right) \! - \!  \sqrt {{{\left( {{a_1}{c_2} \! - \! {a_2}{c_1}} \right)}^2} \!  - \! \left( {{a_1}{b_2} - {a_2}{b_1}} \right)\left( {{b_1} - {b_2}} \right){c_1}} }}{{\left( {{a_1}{b_2} - {a_2}{b_1}} \right)}}
 \end{align} \end{small}and
 \begin{small} \begin{align}\label{eq:p2}
 & {p_2} = \nonumber \\
&\frac{{ \! - \!\left( {{a_1}{c_2}  \! - \!  {a_2}{c_1}} \right) \! + \!  \sqrt {{{\left( {{a_1}{c_2} \! - \! {a_2}{c_1}} \right)}^2} \!  - \! \left( {{a_1}{b_2} - {a_2}{b_1}} \right)\left( {{b_1} - {b_2}} \right){c_1}} }}{{\left( {{a_1}{b_2} - {a_2}{b_1}} \right)}}
 \end{align} \end{small}where
    \begin{small} \begin{align}\label{eq:a2}
{a_1} &= -{\gamma ^2}\left( {\left( {{N_t} - K} \right){\theta _m} - \left( {{N_t} - K} \right){\theta _{b,p}} - K{\theta _{b,q}}} \right){K{\theta _{e,q}}} \\
{b_1} &= \gamma \left( {\left( {{N_t} - K} \right){\theta _m} + \left( {{N_t} - K} \right){\theta _{b,p}} - K{\theta _{b,q}}} \right) \nonumber \\
&\times \left( {\gamma {\theta _{e,q}} + {N_t} - K} \right) + \gamma \left( {\gamma {\theta _{b,q}} + {N_t} - K} \right) \nonumber \\
&\times \left( {\left( {{N_t} - K} \right){\theta _{e,q}} - K{\theta _{e,q}}} \right)  \\
{c_1}  &  = \left( {\gamma {\theta _{b,q}} + {N_t} - K} \right)\left( {\gamma {\theta _{e,q}} + {N_t} - K} \right) \\
{a_{2}}  & = {\gamma ^2}\left( {\left( {{N_t} - K} \right){\theta _{b,p}} - K{\theta _{b,q}}} \right)  \left( {\left( {{N_t} - K} \right){\theta _{e,e}}  - K{\theta _{e,q}}} \right) \\
{b_{2}} & = \gamma \left( {\gamma {\theta _{b,q}} + {N_t} - K} \right)  \left( {\left( {{N_t} - K} \right){\theta _{e,e}}  - K{\theta _{e,q}}} \right) \nonumber \\
 & + \gamma \left( \left( {{N_t} - K} \right) {\theta _{b,p}} - K{\theta _{b,q}} \right) \left( {\gamma {\theta _{e,q}} + {N_t} - K} \right)  \\
 {c_{2}} & = \left( {\gamma {\theta _{b,q}} + {N_t} - K} \right)\left( {\gamma {\theta _{e,q}} + {N_t} - K} \right).
 \end{align} \end{small}

 Then, the optimal power allocation $p$ maximizing the asymptotic achievable secrecy rate in (\ref{eq:asy_sec})
 is given by
 \begin{small} \begin{align}\label{eq:p_opt}
 p^{*} = \left\{ \begin{array}{lll}
 1, &{ {\rm if}}&    {p_1} \notin \left[ {0,1} \right],{p_2} \notin \left[ {0,1} \right], \\
\arg \mathop { \max }\limits_{p \in \left\{ {1,{p_1}} \right\}} {R_{\sec ,\, {\rm asy}}}\left( p \right),  &{ {\rm if}}&    {p_1} \in \left[ {0,1} \right],{p_2} \notin \left[ {0,1} \right], \\
\arg \mathop {\max}\limits_{p \in \left\{ {1,{p_2}} \right\}}  {R_{\sec ,\, {\rm asy}}} (p),  &{ {\rm if}}&   {p_1}
\notin \left[ {0,1} \right],{p_2} \in \left[ {0,1} \right], \\
\arg \mathop {\max}\limits_{p \in \left\{ {1,{p_1},{p_2}} \right\}}  {R_{\sec ,\, {\rm asy}}} (p),  &{ {\rm if}}&   {p_1} \in \left[ {0,1} \right], {p_2} \in \left[ {0,1} \right].
\end{array} \right.
 \end{align} \end{small}

\end{corollary}

\begin{corollary} \label{coro:sec_cond}
 For ${R_{\sec ,\, {\rm asy}}} > 0$, the power allocated to the transmit signal must satisfy\footnote{When ${{a_1} - {a_2}} = 0$, if $b_1 - b_2 > 0$, then secure transmission can be achieved for
any $p$; otherwise, secure transmission can  not be achieved regardless of the value of $p$.}:

$ p >  - \frac{{\left( {{b_1} - {b_2}} \right)}}{{\left( {{a_1} - {a_2}} \right)}}, \, {\rm if} \ \left( {{a_1} - {a_2}} \right) > 0$;

$ p <  - \frac{{\left( {{b_1} - {b_2}} \right)}}{{\left( {{a_1} - {a_2}} \right)}}, \, {\rm if} \  \left( {{a_1} - {a_2}} \right) < 0$.

\end{corollary}

For the special case of single-cell single-user communication ($L = 0, K = 1, m = 1$) with i.i.d. fading
 (${{\bf{R}}^{0}_{01}} = {\beta _B}{{\bf{I}}_{{N_t}}}$ and ${{\bf{R}}_{E}^{0}} = {\beta _E}{{\bf{I}}_{{N_t}}}$, where ${\beta _B}$
 and ${\beta _E}$ denote the path-loss for the desired user and the eavesdropper, respectively.  )
 \cite{Zhou2012TWC}, we have the following result.

\begin{corollary} \label{coro:iid_sec}
$p < {p_{{\rm th},1}}$ has to hold in order to achieve secure transmission in single-cell single-user communication with i.i.d. fading,
where
 \begin{small} \begin{multline}\label{eq:p_th_1}
{p_{{\rm th},\,1}} = 1 + \frac{{\left( {{N_0} + \tau \left( {{P_{01}}{\beta _B} + {P_E}{\beta _E}} \right)} \right)\left( {{P_{01}}\beta _B^2 - {P_E}\beta _E^2} \right)}}{{{P_{01}}
\beta _B^2\gamma {\beta _E}\left( {{N_0} + \tau {P_{01}}{\beta _B}} \right)}} \\
 - \frac{{{P_E}{\beta _E}\left( {{N_0} + \tau {P_E}{\beta _E}} \right)}}{{{P_{01}}{\beta _B}\left( {{N_0} + \tau {P_{01}}{\beta _B}} \right)}}.
\end{multline} \end{small}${p_{{\rm th},\,1}}$
is a decreasing function of $\gamma$. In the high SNR regime, when $\gamma  \to \infty $,
\begin{small} \begin{align}\label{eq:p_th_high}
p_{{\rm th},\,1}  \mathop  \to \limits^{\gamma  \to \infty }  1 - \frac{{{P_E}{\beta _E}\left( {{N_0} + \tau {P_E}{\beta _E}} \right)}}{{{P_{01}}{\beta _B}\left( {{N_0} + \tau {P_{01}}{\beta _B}} \right)}}.
\end{align} \end{small}
\end{corollary}
{\emph{Remark 2:}} We note from (\ref{eq:p_th_1}) that for single-cell single-user communication with i.i.d. fading, to enable secure transmission, the power allocated to the transmit signal has to
decrease with increasing SNR.  In the high SNR regime,  as shown in (\ref{eq:p_th_high}), if the eavesdropper increases the average transmit power $P_E$
in the uplink training phase to make ${p_{{\rm th},\,1}} < 0$, secure transmission cannot be achieved due to the pilot contamination attack.
In this case, increasing the average transmit power of the desired user $P_{01}$ in the uplink training phase is  essential to ensure secure transmission.

\section{Null Space Transmission Design}
In the following proposition, we investigate the
asymptotic achievable secrecy rate under a special condition where the transmit correlation matrices
of the users are orthogonal to the transmit correlation matrices of the eavesdropper.
\begin{prop}\label{prop:sec_orth}
If $\sum\nolimits_{t = 0}^L {P_{tm}} \tr\left( {{\bf{R}}_{tm}^l}{{\bf{R}}_E^{l}}\right) = 0$ for $l = 0,1,\cdots,L$,
then the secrecy rate ${R_{\sec, \, \rm{asy}}}$ is equivalent to the rate without the eavesdropper $R_{\sec, \, \rm{asy}, \, \rm{orth}} =
\log_2(1 + {\rm SINR}_{0m,\,\rm{asy},\,\rm{orth}})$,
where
\begin{small} \begin{align}\label{eq:0m_orth}
{\rm SINR}_{0m,\,{\rm asy},\,\rm{orth}} = \frac{{ p \gamma {\theta _{m,\rm{orth}}}}}{{p\gamma {\theta _{b,p,\rm{orth}}} +  q\gamma {\theta _{b,q,\rm{orth}}} + 1}}
\end{align} \end{small}with
\begin{small} \begin{align}
{\theta _{m,\rm{orth}}} &= \tr\left( {\widehat {\bf{R}}_{0m,\rm{orth}}^0} \right) \nonumber \\
&+ \tr{\left( {\widehat {\bf{R}}_{0m,\rm{orth}}^0} \right)^{ - 1}}\!\tr\left(\! {\left( \! {{\bf{R}}_{0m}^0 - \widehat {\bf{R}}_{0m,\rm{orth}}^0} \right)\widehat {\bf{R}}_{0m,\rm{orth}}^0} \right) \\
{\theta _{b,p,\rm{orth}}} &  =   \sum\limits_{l = 0}^L {\sum\limits_{k = 1,k \ne m}^K {\tr{{\left( {\widehat {\bf{R}}_{lk}^l} \right)
}^{ - 1}}\tr\left( {{\bf{R}}_{0m}^l\widehat {\bf{R}}_{lk}^l} \right)}}
\nonumber \\ &  +  \sum\limits_{l = 1}^L { \!\tr{{\left(\! {\widehat {\bf{R}}_{lm,{\rm orth}}^l} \!\right)}^{ - 1}} \! \Lambda _{0m,{\rm orth}}^l}  \\
{\theta _{b,q,\rm{orth}}} &  =  \sum\limits_{l = 0}^L {\tr\left( {{\bf{R}}_{0m}^l} \right)} \! - \! \sum\limits_{l = 0}^L \sum\limits_{k = 1,k \ne m}^K {\tr{{\left( {\widehat {\bf{R}}_{lk}^l} \right)}^{ - 1}}\tr\left( {{\bf{R}}_{0m}^l\widehat {\bf{R}}_{lk}^l} \right)} \nonumber \\
&  - \sum\limits_{l = 0}^L {\tr{{\left( {\widehat {\bf{R}}_{lm,\rm{orth}}^l} \right)}^{ - 1}}\Lambda _{0m,\rm{orth}}^l}
\end{align} \end{small}
\begin{small} \begin{align}
\hspace{-0.7cm}  \Lambda _{0m,\rm{orth}}^l &= {\tau ^2}{P_{0m}}{\left| {\tr\left( {{\bf{C}}_{lm,\rm{orth}}^l{\bf{R}}_{0m}^l} \right)} \right|^2} \nonumber  \\
& + {\tau ^2}\sum\limits_{t = 1}^L {{P_{tm}}\tr\left( {{\bf{R}}_{0m}^l{\bf{C}}_{lm,\rm{orth}}^l{\bf{R}}_{tm}^l
{{\left( {{\bf{C}}_{lm,\rm{orth}}^l} \right)}^H}} \right)} \nonumber  \\
&  +   \tau {N_0}\tr\left( {{\bf{R}}_{0m}^l{\bf{C}}_{lm,\rm{orth}}^l{{\left( {{\bf{C}}_{lm,\rm{orth}}^l} \right)}^H}} \right)
\end{align} \end{small}
\begin{small} \begin{align}
\hspace{-1.5cm} {\bf{C}}_{lm,\rm{orth}}^l    = \sqrt {{P_{lm}}} {\bf{R}}_{lm}^l{\left( {{N_0}{{\bf{I}}_{{N_t} }}
 \! + \! \tau {\sum\limits_{t = 0}^L {{P_{tm}}{\bf{R}}_{tm}^l} } } \right)^{ - 1}}
\label{eq:Cllm} \end{align}\end{small}
\begin{small} \begin{align}
\hspace{-0.2cm} \widehat {\bf{R}}_{lm,\rm{orth}}^l   = {P_{lm}}\tau {\bf{R}}_{lm}^l{\left( {{N_0}{{\bf{I}}_{{N_t}}} + \tau \sum\limits_{t = 0}^L {{P_{tm}}{\bf{R}}_{tm}^l}  } \right)^{ - 1}}{\bf{R}}_{lm}^l.
\end{align} \end{small}

\begin{proof}
The proposition can be proved by applying the matrix inversion lemma in (\ref{eq:R0m0_est}) and (\ref{eq:Rnmn_est}),
using the fact ${\bf{R}}_E^0 {\bf{C}}_{0m}^0 = \mathbf{0}$, and performing some matrix transformation
operations. Due to space limitation, details of the proof are omitted
here.
\end{proof}
\end{prop}

Proposition \ref{prop:sec_orth} shows that when
the channels of the eavesdropper  and the users are statistically orthogonal, the impact of the pilot contamination attack disappears.
It is known that for many massive MIMO scenarios, the transmit correlation matrices of the channels are low rank \cite{Yin2013JSAC,Adhikary2013TIT}. As a result, inspired by Proposition \ref{prop:sec_orth},
we propose a null space transmission design  along the correlation matrix of the eavesdropper's channel $\mathbf{h}_{E}^{0}$ as follows.

In the uplink training phase, we can rewrite (\ref{eq:y_0_multi}) as
 \begin{small} \begin{align} \label{eq:y_0_multi_null_0}
 {{\bf{Y}}_0} & =   \sum\limits_{k = 1}^K  \! {\sqrt {{P_{0k}}}{\bf{h}}_{0k}^0} {\boldsymbol{\omega }}^T_{k}  +   \sum\limits_{l = 1}^L  \sum\limits_{k = 1}^K  {\sqrt {{P_{lk}}} {\bf{h}}_{lk}^0} {\boldsymbol{\omega }}_{k}^T
+ \sqrt {{P_E}}{\bf{h}}_E^0 {\boldsymbol{\omega }}_{m}^T + {\bf{N}}
\end{align} \end{small}where
${\bf{y}}_0 = {\rm vec}\left( {\bf{Y}}_0 \right)$ and $\mathbf{n} = {\rm vec}\left( {\bf{N}} \right)$.
Assume the rank of $\mathbf{R}_E^0$ is $N$.
Let us now construct a matrix ${{\bf{V}}_E^0} \in {\mathbb{C}^{{N_t} \times M}}$,
whose $M$ columns are the $M$ eigenvectors which correspond to the zero eigenvalues of $\mathbf{R}_E^0$, where $M = N_t - N$.
In the uplink training phase, we multiply ${\bf{Y}}_{0}$  with ${{\bf{V}}_E^0}$ to obtain
 \begin{small} \begin{multline} \label{eq:y_0_multi_null}
 \left({{\bf{V}}_E^0}\right)^H  {{\bf{Y}}_0}  =   \sum\limits_{k = 1}^K  \! {\sqrt {{P_{0k}}}   \left({{\bf{V}}_E^0}\right)^H{\bf{h}}_{0k}^0} {\boldsymbol{\omega }}^T_{k} +   \sum\limits_{l = 1}^L  \sum\limits_{k = 1}^K  \sqrt {{P_{lk}}}  \\
 \times \left({{\bf{V}}_E^0}\right)^H{\bf{h}}_{lk}^0 {\boldsymbol{\omega }}_{k}^T  + \sqrt {{P_E}} \left({{\bf{V}}_E^0}\right)^H
 {\bf{h}}_E^0 {\boldsymbol{\omega }}_{m}^T +  \left({{\bf{V}}_E^0}\right)^H {\bf{N}}
\end{multline} \end{small}and
 \begin{small} \begin{align}\label{eq:y_uplink_null}
 {{\bf{y}}_{0,\rm null}}& = {\rm vec}\left( { \left({{\bf{V}}_E^0}\right)^H  {\bf{Y}}_0} \right)= \sum\limits_{k = 1}^K  \! {\sqrt {{P_{0k}}} \left( {{{\boldsymbol{\omega }}_{k}} \otimes {{\bf{I}}_{{N_t}}}} \right){\bf{h}}_{0k,\rm{null}}^0} \nonumber \\
  &  +  \sum\limits_{l = 1}^L  \! {\sum\limits_{k = 1}^K  \! {\sqrt {{P_{lk}}} \left( {{{\boldsymbol{\omega }}_{k}} \otimes {{\bf{I}}_{{N_t}}}} \right){\bf{h}}_{lk,\rm{null}}^0} } \nonumber \\
 & + \sqrt {{P_E}} \left( {{{\boldsymbol{\omega }}_{m}} \otimes {{\bf{I}}_{{N_t}}}} \right) \left({{\bf{V}}_E^0}\right)^H {\bf{h}}_{E}^0 + {\bf{\widetilde{n}}}_{\rm null}
\end{align} \end{small}where
${\bf{h}}_{lk,\rm{null}}^0 = \left({{\bf{V}}_E^0}\right)^H {\bf{h}}_{lk}^0  \sim \mathcal{CN}\left( {0,{{\bf{R}}_{lk,\rm{null}}^0}} \right)$,
 ${{\bf{R}}_{lk,\rm{null}}^0} = \left({{\bf{V}}_E^0}\right)^H {{\bf{R}}_{lk}^0} {{\bf{V}}_E^0}  $,
$l = 0,1,\cdots,L$, $k = 1,2,\cdots,K$, and $\widetilde {\bf{n}}_{\rm null} \sim \mathcal{CN}\left( {0, \tau N_0 \mathbf{I}_M} \right)$.
Then, we estimate $\mathbf{h}_{0m,\rm{null}}^0$ as
 \begin{small} \begin{multline} \label{eq:mmse_h0m_null}
{\widehat {\bf{h}}_{0m,\rm{null}}^0} = \\
\sqrt {{P_{0m}}} {\bf{R}}_{0m,\rm{null}}^0{\left( \! {{N_0}{{\bf{I}}_{{M}}}  +  \tau  {\sum\limits_{t = 0}^L
{{P_{tm}}{\bf{R}}_{tm,\rm{null}}^0} }} \right)^{ - 1}}{\widetilde {\bf{y}}_{0m,\rm{null}}},
\end{multline} \end{small}where
 \begin{small} \begin{multline} \label{eq:y_0m_null}
{\widetilde {\bf{y}}_{0m,\rm{null}}} = \sqrt {{P_{0m}}} \tau {\bf{h}}_{0m,\rm{null}}^0 + \sum\limits_{t = 1}^L {\sqrt {{P_{tm}}} \tau {\bf{h}}_{tm,\rm{null}}^0} \\
 + \sqrt {{P_E}} \tau  \left({{\bf{V}}_E^0}\right)^H {\bf{h}}_E^0 + {\left( {{{\boldsymbol{\omega }}_m} \otimes {{\bf{I}}_{{M}}}} \right)^H}{\bf{\widetilde{n}}_{\rm null}}.
\end{multline} \end{small}Similarly,
we have  ${\bf{h}}_{0m,\rm{null}}^0 = {\bf{\widehat h}}^0_{0m,\rm{null}} + {\bf{e}}_{0m,\rm{null}}^0$, where
${\bf{\widehat h}}^0_{0m} \sim \mathcal{CN} \left({\mathbf{0}_{N_t}, {\widehat {\bf{R}}_{0m,\rm{null}}^0} }\right)$ and ${\bf{e}}^0_{0m,\rm{null}} \sim \mathcal{CN} \left({\mathbf{0}_{N_t}, {{\bf{R}}_{0m,\rm{null}}^0} - {\widehat {\bf{R}}_{0m,\rm{null}}^0} }\right)$, with
 \begin{small} \begin{multline} \label{eq:R0m0_est_null}
\widehat {\bf{R}}_{0m,\rm{null}}^0 = \\
{P_{0m}}\tau {\bf{R}}_{0m,\rm{null}}^0{\left( {N_0}{{\bf{I}}_{{M}}}  +  \tau  {\sum\limits_{t = 0}^L {{P_{tm}}{\bf{R}}_{tm,\rm{null}}^0}} \right)^{ - 1}}
{\bf{R}}_{0m,\rm{null}}^0.
\end{multline} \end{small}

In the downlink data transmission phase, we employ the same transmission scheme as in Section II-B, but replace $\mathbf{\bf{w}}_{0m}$
with ${{\bf{w}}_{0m,\rm{null}}} =  \mathbf{V}_E^0 \frac{{\widehat {\bf{h}}_{0m,\rm{null}}^0}}{{\left\| {\widehat {\bf{h}}_{0m,\rm{null}}^0} \right\|}}$,
and set $p = 1/K$, $q = 0$. We refer to this transmission design as ``Null Space Design". Following the same approach as in Appendix A, we can show
that the asymptotic achievable secrecy rate of the ``Null Space Design" is still
given by (\ref{eq:asy_sec}) after replacing
${\bf{R}}_{0m}$ and $\widehat {\bf{R}}_{0m}^0$  with ${\bf{R}}_{0m,\rm{null}}$ and $\widehat {\bf{R}}_{0m,\rm{null}}^0$
in $\theta_m$, respectively, and setting $\mathbf{R}_E^{0} = \mathbf{0}$ in (\ref{eq:asy_sec}). The other terms in (\ref{eq:asy_sec})
remain unchanged.

{\emph{Remark 3:}} The proposed null space design transmits the signal
along the orthogonal subspace $\mathbf{V}_E^{0}$.  As a result, the
performance of the null space design depends on the rank of $\mathbf{V}_E^{0}$.
For instance, for the extreme case of i.i.d. fading,  $\mathbf{V}_E^{0}$ does not exist
and hence  the null space design is not applicable. In practice, the null space design will perform
better in highly correlated channels, under strong pilot contamination attack, and in the high SNR
regime since it can effectively degrade the eavesdropper's performance in these scenarios.
Matched filter precoding and AN generation, in contrast,
will perform better in weakly correlated channels, under weak pilot contamination attack,
and in the low SNR regime.  Based on the
expression in (\ref{eq:asy_sec}), an analytical criterion can be obtained to determine
whether matched filter precoding with AN generation or the null space design
is preferable for a given channel and eavesdropper parameters.

\section{Numerical Results}
In this section, we provide numerical results to evaluate the secrecy performance of the considered massive MIMO system with
an active eavesdropper. We consider a system where a $N_t = 128$ uniform linear array is employed at the base station, the antenna
spacing is  half a wavelength, and the angle of arrival (AoA) interval is $\mathcal{A} = [-\pi/2,\pi/2]$. We use the truncated Laplacian distribution
to model the channel power angle spectrum as \cite{Cho2010}
\begin{small} \begin{align} \label{eq:p_theta}
p\left( \theta  \right) = \frac{1}{{\sqrt 2 \sigma \left( {1 - {e^{ - \sqrt 2 \pi /\sigma }}} \right)}}{e^{\frac{{ - \sqrt 2 \left\| {\theta  - \overline \theta  } \right\|}}{\sigma }}},
\end{align} \end{small}where
$\sigma$ and $\overline\theta$ denote the angular spread (AS) and the mean  AoA of the channel, respectively.
We assume that the ASs $\sigma$  in (\ref{eq:p_theta}) for the channels of all  users and the eavesdropper are identical and we
set $\sigma = \pi/2$.
The channel covariance matrices of all users and the eavesdropper are generated based on \cite[Eq. (3.14)]{Cho2010}. For the channel between the user
and the BS in its own cell and the channel between the user
and the BSs in the other cells, we impose a channel power normalization
to make the trace of the channel covariance matrices equal to $N_t$ and $\rho N_t$, respectively, and
set $\rho = 0.1$. The asymptotic secrecy rate is computed from
Proposition  \ref{prop:sec_rate_mul} and the exact secrecy rate is obtained by Monte Carlo
simulation. We set $L = 3$, $K = 5$, $P_{lk} =1$, $k = 1,2,\cdots,K$, $l = 0,1,\cdots,L$, $\tau = 10$, and $N_0 = 1$. The mean channel AoAs $\overline \theta$ of all  users and the eavesdropper in (\ref{eq:p_theta}) are generated at random and
the channel AoAs $ \theta$ of all users and the eavesdropper
are distributed within the angle interval $\left[-\pi,  \pi\right]$ based on (\ref{eq:p_theta}).

Figure \ref{Sec_Asy_SNR_Multi} shows the secrecy rate performance vs. the
SNR $\gamma$ for matched filter precoding and AN
generation and different $p$. We observe from Figure \ref{Sec_Asy_SNR_Multi} that the asymptotic secrecy rate
in Proposition \ref{prop:sec_rate_mul} provides a good estimate
for the exact secrecy rate. Also, we observe from Figure \ref{Sec_Asy_SNR_Multi} that in the low SNR regime,
allocating more power (larger $p$) to the transmit signal leads to a higher secrecy rate. However, as the SNR increases,
the secrecy rate may be small if the transmit signal power is high, i.e., if the AN power is small.
For example, for ${\rm SNR} = 2$ dB,  we find from  Corollary \ref{coro:sec_cond}
that $p < 0.15$ is a necessary condition to guarantee reliable communication.
Thus, for $p = 0.18$, a positive secrecy rate cannot be achieved, as confirmed by Figure \ref{Sec_Asy_SNR_Multi}.

Figure \ref{Sec_Asy_SNR_Multi_Unified} shows the exact secrecy rate performance vs. the SNR for different $P_E$ and different designs.
For matched filter precoding with AN generation, we
obtain the optimal power allocation $p$ and $q$ based on Corollary \ref{prop:sec_rate_mul}.
For naive matched filter precoding, we set $p = 1/K$ and do not generate
AN.  From Figure \ref{Sec_Asy_SNR_Multi_Unified} we make the following observations:
1) Because of the pilot contamination attack, even if the transmitter is equipped with a large number of antennas,
naive matched filter precoding can not achieve a positive secrecy rate for moderate-to-high SNR.
2) For $P_E = 1$,  matched filter precoding with AN
generation performs better in the low-to-moderate SNR regime and
the null space design performs better in the moderate-to-high SNR regime, see Remark 3.
3) When the eavesdropper increases the
power of the pilot contamination attack, this results in a serious secrecy rate loss
for matched filter precoding with AN generation, but  has barely any visible impact on the null space design.
This is because the null space design can eliminate the
impact of pilot contamination caused by the active eavesdropper  as suggest by  (\ref{eq:mmse_h0m_null})
and significantly degrade the achievable
rate of the active eavesdropper.

\begin{figure}[!t]
\centering
\includegraphics[width=0.4\textwidth]{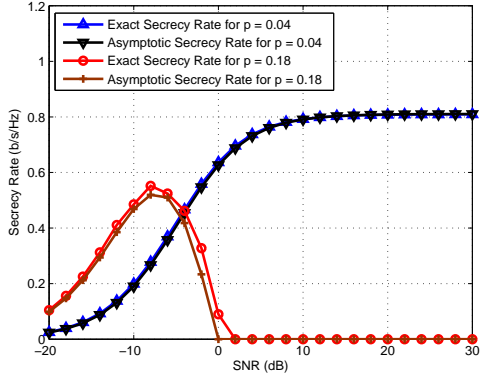}
\caption {\space\space  Secrecy rate vs. the SNR for different $p$}
\label{Sec_Asy_SNR_Multi}
\end{figure}

\begin{figure}[!t]
\centering
\includegraphics[width=0.4\textwidth]{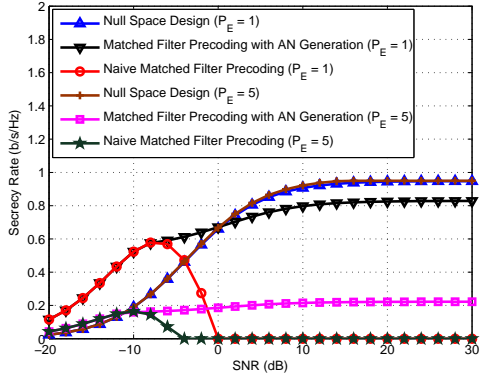}
\caption {\space\space  Exact secrecy rate vs. the SNR for different $P_E$ and different precoding designs.}
\label{Sec_Asy_SNR_Multi_Unified}
\end{figure}

\section{Conclusions}
In this paper, we have studied the transmission design for multi-cell multi-user massive
MIMO systems in the presence of an active eavesdropper. Assuming
matched filter precoding and AN generation, we obtained an asymptotic
achievable secrecy rate expression for the case of a pilot contamination attack
 when the number of transmit antennas tends to infinity.
We obtained a closed-form expression
for the optimal power allocation policy for the transmit signal and the AN.
Moreover, we proved that the impact of the active eavesdropper is completely eliminated when
the transmit correlation matrices of the users and the eavesdropper are orthogonal.
This motivates  the development of  a null space  design that exploits
the low rank property of the transmit correlation matrices of massive MIMO channels.
Monte Carlo simulation results showed that the derived analytical results are accurate
and confirmed the effectiveness of the proposed transmission schemes for combating
the pilot contamination attack.

\appendices

\section{Proof of Proposition \ref{prop:sec_rate_mul}}\label{proof:prop:sec_rate_mul}
First, we calculate ${\rm SINR}_{0m}$ in (\ref{eq:SINR_0m}). For the numerator of (\ref{eq:SINR_0m}), based on \cite[Corollary 1]{Evans2000TIT},  we have
 \begin{small} \begin{align} \label{eq:h0mw0m_2_app}
& \frac{1}{{{N_t}}} \left[ {{{\left| {{{\left( {{\bf{h}}_{0m}^0} \right)}^H}{{\bf{w}}_{0m}}} \right|}^2}} \right] \nonumber \\
& \mathop  \to \limits^{{N_t} \to \infty} \frac{1}{{{N_t}}}  \left| {\widehat {\bf{h}}_{0m}^0} \right|^2
+ \frac{1}{{{N_t}}}\left[ {\frac{{{{\left( {\widehat {\bf{h}}_{0m}^0} \right)}^H}\left( {{\bf{R}}_{0m}^0 - \widehat {\bf{R}}_{0m}^0} \right)\widehat {\bf{h}}_{0m}^0}}{{{{\left\| {\widehat {\bf{h}}_{0m}^0} \right\|}^2}}}} \right] \\
& \mathop  \to \limits^{{N_t} \to \infty} \frac{1}{{{N_t}}}\tr\left( {\widehat {\bf{R}}_{0m}^0} \right)
+ \frac{1}{{{N_t}}} \frac{{\tr\left({\left({{\bf{R}}_{0m}^0 - \widehat{\bf{R}}_{0m}^0} \right) \widehat{\bf{R}}_{0m}^0} \right)}}{{\tr\left( {\widehat {\bf{R}}_{0m}^0} \right)}}.
\end{align} \end{small}

For the denominator of (\ref{eq:SINR_0m}), we have
\begin{small} \begin{align} \label{eq:h0mw0k_app}
\frac{1}{{{N_t}}} {{{\left| {{{\left( {{\bf{h}}_{0m}^0} \right)}^H}{{\bf{w}}_{0k}}} \right|}^2}} & =
{\frac{{\left( {{\bf{h}}_{0m}^0} \right)}^H \widehat {\bf{h}}_{0k}^0 {\left( {\widehat {\bf{h}}_{0k}^0} \right)}^H
{\left( {{\bf{h}}_{0m}^0} \right)}
}{{{{\left( {\widehat {\bf{h}}_{0k}^0} \right)}^H}\widehat {\bf{h}}_{0k}^0}}} \\
& \mathop  \to \limits^{{N_t} \to \infty }   {\frac{{{{\left( {\widehat {\bf{h}}_{0k}^0} \right)}^H}{\bf{R}}_{0m}^0\widehat {\bf{h}}_{0k}^0}}{{{{\left( {\widehat {\bf{h}}_{0k}^0} \right)}^H}\widehat {\bf{h}}_{0k}^0}}}  \\
& \mathop  \to \limits^{{N_t} \to \infty } \frac{1}{{{N_t}}} \frac{{\tr\left( {{\bf{R}}_{0m}^0\widehat {\bf{R}}_{0k}^0} \right)}}{{\tr\left( {\widehat {\bf{R}}_{0k}^0} \right)}}. \label{eq:h0mw0k_app_2}
\end{align} \end{small}Also, performing some simplifications, we have
 \begin{small} \begin{align} \label{eq:h0mUull0_app}
& \frac{1}{{{N_t}}}\left[ {{{\left\| {{{\left( {{\bf{h}}_{0m}^0} \right)}^H}{{\bf{U}}_{{\rm null},\,0}}} \right\|}^2}} \right] \mathop  \to \limits^{{N_t} \to \infty }
 \frac{1}{{{N_t}}} {{\bf{R}}_{0m}^0} \nonumber \\
&  - \frac{1}{{{N_t}}}\sum\limits_{k = 1,k \ne m}^K {\tr{{\left( {\widehat {\bf{R}}_{0k}^0} \right)}^{ - 1}} {{{\left( {{\bf{h}}_{0m}^0} \right)}^H}\left( {\widehat {\bf{h}}_{0k}^0} \right){{\left( {\widehat {\bf{h}}_{0k}^0} \right)}^H}{\bf{h}}_{0m}^0}} \nonumber \\
&  - \frac{1}{{{N_t}}}\tr{\left( {\widehat {\bf{R}}_{0m}^0} \right)^{ - 1}} {{{\left( {{\bf{h}}_{0m}^0} \right)}^H}\left( {\widehat {\bf{h}}_{0m}^0} \right){{\left( {\widehat {\bf{h}}_{0m}^0} \right)}^H}{\bf{h}}_{0m}^0}.
\end{align} \end{small}

For $m \ne k$, ${\bf{h}}_{0m}^0$ is independent of $\widehat {\bf{h}}_{0k}^0$. Hence, the
asymptotic expression for $ \frac{1}{{{N_t}}} {{{\left( {{\bf{h}}_{0m}^0} \right)}^H}\widehat {\bf{h}}_{0k}^0{{\left( {\widehat {\bf{h}}_{0k}^0} \right)}^H}{\bf{h}}_{0m}^0}$
is given by the numerator of (\ref{eq:h0mw0k_app_2}).

For $m = k$, based on (\ref{eq:mmse_h0m}), we have
 \begin{small} \begin{align} \label{eq:h0m_app}
\widehat {\bf{h}}_{0m}^0 & = \sqrt {{P_{0m}}} \tau {\bf{C}}_{0m}^0{\bf{h}}_{0m}^0 + {\bf{C}}_{0m}^0\sum\limits_{l = 1}^L {\sqrt {{P_{lm}}} \tau {\bf{h}}_{lm}^0} \nonumber \\
& + {\bf{C}}_{0m}^0\sqrt {{P_E}} \tau {\bf{h}}_E^0 + {\bf{C}}_{0m}^0  {{\bf{\Omega }}_m} {\bf{n}},
\end{align} \end{small}where
 ${\bf{C}}_{0m}^0$ is defined in (\ref{eq:Cllm}) and ${{\bf{\Omega }}_m} =  {\left( {{{\boldsymbol{\omega }}_m} \otimes {{\bf{I}}_{{N_t}}}} \right)^H}$.

Based on (\ref{eq:h0m_app}), we obtain
 \begin{small} \begin{align} \label{eq:h0mhath0m_app}
& {\left( {{\bf{h}}_{0m}^0} \right)^H}\widehat {\bf{h}}_{0m}^0{\left( {\widehat {\bf{h}}_{0m}^0} \right)^H}{\bf{h}}_{0m}^0 = {\left( {{\bf{h}}_{0m}^0} \right)^H}{\bf{C}}_{0m}^0{{\bf{\Omega }}_m}{\bf{{n}}}{\left( {\widehat {\bf{h}}_{0m}^0} \right)^H}{\bf{h}}_{0m}^0 \nonumber \\
& + {\tau ^2}{\left( {{\bf{h}}_{0m}^0} \right)^H}{\bf{C}}_{0m}^0\sum\limits_{t = 0}^L {\sum\limits_{s = 0}^L {\sqrt {{P_{tm}}} \sqrt {{P_{sm}}} {\bf{h}}_{tm}^0} } {\left( {{\bf{h}}_{sm}^0} \right)^H}{\left( {{\bf{C}}_{0m}^0} \right)^H}{\bf{h}}_{0m}^0 \nonumber \\
& + {\tau ^2}\sqrt {{P_E}} {\left( {{\bf{h}}_{0m}^0} \right)^H}{\bf{C}}_{0m}^0\sum\limits_{t = 0}^L {\sqrt {{P_{tm}}} {\bf{h}}_{tm}^0{{\left( {{\bf{h}}_E^0} \right)}^H}} {\left( {{\bf{C}}_{0m}^0} \right)^H}{\bf{h}}_{0m}^0  \nonumber \\
& + \tau {\left( {{\bf{h}}_{0m}^0} \right)^H}{\bf{C}}_{0m}^0\sum\limits_{t = 0}^L {\sqrt {{P_{tm}}} {\bf{h}}_{tm}^0{{ {\bf{n}}}^H  {{\bf{\Omega }}_m^H}    }} {\left( {{\bf{C}}_{0m}^0} \right)^H}{\bf{h}}_{0m}^0 \nonumber \\
& + \tau \sqrt {{P_E}} {\left( {{\bf{h}}_{0m}^0} \right)^H}{\bf{C}}_{0m}^0{\bf{h}}_E^l{\left( {\widehat {\bf{h}}_{0m}^0} \right)^H}{\bf{h}}_{0m}^0
\end{align} \end{small}When
$N_t \rightarrow \infty$, based on (\ref{eq:h0mhath0m_app}) and \cite[Corollary 1]{Evans2000TIT},  we have
\begin{small} \begin{align}
& \frac{1}{{{N_t}}}  {\left( {{\bf{h}}_{0m}^0} \right)^H}\widehat {\bf{h}}_{0m}^0{\left( {\widehat {\bf{h}}_{0m}^0} \right)^H}{\bf{h}}_{0m}^0 \nonumber \\
&\mathop  \to \limits^{{N_t} \to \infty } \frac{1}{{{N_t}}}  {\tau ^2}{\left( {{\bf{h}}_{0m}^0} \right)^H}{\bf{C}}_{0m}^0\sum\limits_{t = 0}^L {{P_{tm}}{\bf{h}}_{tm}^0{{\left( {{\bf{h}}_{tm}^0} \right)}^H}{{\left( {{\bf{C}}_{0m}^0} \right)}^H}{\bf{h}}_{0m}^0} \nonumber  \\
& + \frac{1}{{{N_t}}} {\tau ^2}{P_E}{\left( {{\bf{h}}_{0m}^0} \right)^H}{\bf{C}}_{0m}^0{\bf{h}}_E^0{\left( {{\bf{h}}_E^0} \right)^H}{\left( {{\bf{C}}_{0m}^0} \right)^H}{\bf{h}}_{0m}^0 \nonumber \\
&  + \frac{1}{{{N_t}}} {\left( {{\bf{h}}_{0m}^0} \right)^H}{\bf{C}}_{0m}^0{{\bf{\Omega }}_m}{\bf{n}}{\left( {{\bf{C}}_{0m}^0{{\bf{\Omega }}_m}{\bf{n}}} \right)^H}{\bf{h}}_{0m}^0  \nonumber  \\
& = \frac{1}{{{N_t}}}  {\tau ^2}{P_{0m}}{\left( {{\bf{h}}_{0m}^0} \right)^H}{\bf{C}}_{0m}^0{\bf{h}}_{0m}^0{\left( {{\bf{h}}_{0m}^0} \right)^H}{\left( {{\bf{C}}_{0m}^0} \right)^H}{\bf{h}}_{0m}^0 \nonumber \\
& + \frac{1}{{{N_t}}} {\tau ^2}{\left( {{\bf{h}}_{0m}^0} \right)^H}{\bf{C}}_{0m}^0\sum\limits_{t = 1}^L {{P_{tm}}{\bf{h}}_{tm}^0{{\left( {{\bf{h}}_{tm}^0} \right)}^H}{{\left( {{\bf{C}}_{0m}^0} \right)}^H}{\bf{h}}_{0m}^0} \nonumber \\
& + \frac{1}{{{N_t}}} {\tau ^2}{P_E}{\left( {{\bf{h}}_{0m}^0} \right)^H}{\bf{C}}_{0m}^0{\bf{h}}_E^0{\left( {{\bf{h}}_E^0} \right)^H}{\left( {{\bf{C}}_{0m}^0} \right)^H}{\bf{h}}_{0m}^0 \nonumber \\
 & + \frac{1}{{{N_t}}} {\left( {{\bf{h}}_{0m}^0} \right)^H}{\bf{C}}_{0m}^0{{\bf{\Omega }}_m}{\bf{n}}{\left( {{\bf{C}}_{0m}^0{{\bf{\Omega }}_m}{\bf{n}}} \right)^H}{\bf{h}}_{0m}^0 \mathop  \to \limits^{{N_t} \to \infty }  \frac{1}{{{N_t}}}  \Lambda _{0m}^0. \label{eq:h0mhath0m_2_2_app}
\end{align} \end{small}

Also, for $m \ne k$, ${\bf{h}}_{0m}^l$ is independent of $\widehat {\bf{h}}_{lk}^l$,
and we obtain in (\ref{eq:SINR_0m_b})
\begin{small} \begin{align} \label{eq:h0mlwlk_app}
\frac{1}{{{N_t}}} {{{\left| {{{\left( {{\bf{h}}_{0m}^l} \right)}^H}{{\bf{w}}_{lk}}} \right|}^2}} & = \frac{1}{{{N_t}}} {{{\left( {{\bf{h}}_{0m}^l} \right)}^H}\frac{{\widehat {\bf{h}}_{lk}^l}}{{\left| {\widehat {\bf{h}}_{lk}^l} \right|}}\frac{{{{\left( {\widehat {\bf{h}}_{lk}^l} \right)}^H}}}{{\left| {\widehat {\bf{h}}_{lk}^l} \right|}}{\bf{h}}_{0m}^l} \\
&\mathop  \to \limits^{{N_t} \to \infty } \frac{{{{\left( {\widehat {\bf{h}}_{lk}^l} \right)}^H}{\bf{R}}_{0m}^l\widehat {\bf{h}}_{lk}^l}}{{{{\left| {\widehat {\bf{h}}_{lk}^l} \right|}^2}}} \mathop  \to \limits^{{N_t} \to \infty } \frac{{\tr\left( {{\bf{R}}_{0m}^l\widehat {\bf{R}}_{lk}^l} \right)}}{{\tr\left( {\widehat {\bf{R}}_{lk}^l} \right)}}.
\end{align} \end{small}

For $m = k$, similar to (\ref{eq:h0mhath0m_2_2_app}), we have
\begin{small} \begin{align} \label{eq:h0mhath0m_3_app}
\frac{1}{{{N_t}}}  {{{\left| {{{\left( {{\bf{h}}_{0m}^l} \right)}^H}{{\bf{w}}_{lm}}} \right|}^2}} &  = \frac{1}{N_t}\frac{\left( {{\bf{h}}_{0m}^l} \right)^H  \widehat {\bf{h}}_{lm}^l  {\left( {\widehat {\bf{h}}_{lm}^l} \right)^H}\! {\bf{h}}_{0m}^l}{\left| {\widehat {\bf{h}}_{lm}^l} \right|^2}   \\
& \mathop  \to \limits^{{N_t}\to \infty  }  \frac{1}{{{N_t}}}  \frac{\Lambda _{0m}^l}{\tr\left( {\widehat {\bf{R}}_{lm}^l} \right)}.
\end{align} \end{small}

Next, we simplify
\begin{small} \begin{align} \label{eq:h0m_null_app}
&  {{{\left( {{\bf{h}}_{0m}^l} \right)}^H}{{\bf{U}}_{{\rm null},\,l}}{{ {{{\bf{U}}_{{\rm null},\,l}^H}} }}{\bf{h}}_{0m}^l}  = \tr\left( {{\bf{R}}_{0m}^l} \right) - \nonumber \\
&  \sum\limits_{k = 1,k \ne m}^K {\tr{{\left( {\widehat {\bf{R}}_{lk}^l} \right)}^{ - 1}} {{{\left( {{\bf{h}}_{0m}^l} \right)}^H}\left( {\widehat {\bf{h}}_{lk}^l} \right){{\left( {\widehat {\bf{h}}_{lk}^l} \right)}^H}{\bf{h}}_{0m}^l} } \nonumber \\
& - \tr{\left( {\widehat {\bf{R}}_{lm}^l} \right)^{ - 1}} {{{\left( {{\bf{h}}_{0m}^l} \right)}^H}\left( {\widehat {\bf{h}}_{lm}^l} \right){{\left( {\widehat {\bf{h}}_{lm}^l} \right)}^H}{\bf{h}}_{0m}^l}.
\end{align} \end{small}

Following a similar approach as was used to obtain (\ref{eq:h0mw0k_app_2}) and (\ref{eq:h0mhath0m_2_2_app}), we obtain
\begin{small} \begin{align} \label{eq:h0mhlk}
& \frac{1}{{{N_t}}} {\left( {{\bf{h}}_{0m}^l} \right)^H}\left( {\widehat {\bf{h}}_{lk}^l} \right){\left( {\widehat {\bf{h}}_{lk}^l} \right)^H}{\bf{h}}_{0m}^l  \mathop  \to \limits^{{N_t}\to \infty  } \tr\left( {{\bf{R}}_{0m}^l\widehat {\bf{R}}_{lk}^l} \right)  \\
& \frac{1}{{{N_t}}} {{{\left( {{\bf{h}}_{0m}^l} \right)}^H}\left( {\widehat {\bf{h}}_{lm}^l} \right){{\left( {\widehat {\bf{h}}_{lm}^l} \right)}^H}{\bf{h}}_{0m}^l}   \mathop  \to \limits^{{N_t}\to \infty}
\Lambda _{0m}^l. \label{eq:h0mhlk_2}
\end{align} \end{small}

By substituting (\ref{eq:h0mw0m_2_app})--(\ref{eq:h0mhlk_2}) into (\ref{eq:SINR_0m}), we obtain the expression for ${\rm SINR}_{0m,\,\rm{asy}}$
in (\ref{eq:b}). Next, we simplify ${\rm SINR}_{{\rm eve}}$ in (\ref{eq:SINR_E_mul}).  First, we consider
\begin{small} \begin{align} \label{eq:hE0w0m_app}
{\left( {{\bf{h}}_E^0} \right)^H}{{\bf{w}}_{0m}}{ {{{\bf{w}}_{0m}^H}}}{\bf{h}}_E^0 = \frac{{{{\left( {{\bf{h}}_E^0} \right)}^H}\widehat {\bf{h}}_{0m}^0{{\left( {\widehat {\bf{h}}_{0m}^0} \right)}^H}{\bf{h}}_E^0}}{{{{\left| {\widehat {\bf{h}}_{0m}^0} \right|}^2}}}.
\end{align} \end{small}

When $N_t \rightarrow \infty$, we have
\begin{small} \begin{align} \label{eq:hEh0m_app}
&\frac{1}{{{N_t}}}{\left( {{\bf{h}}_E^0} \right)^H}\widehat {\bf{h}}_{0m}^0{\left( {\widehat {\bf{h}}_{0m}^0} \right)^H}{\bf{h}}_E^0  \mathop  \to \limits^{{N_t}\to \infty} \frac{1}{{{N_t}}}\Lambda _E^0 \\
& \frac{1}{{{N_t}}}{\left| {\widehat {\bf{h}}_{0m}^0} \right|^2} \mathop  \to \limits^{{N_t}\to \infty} \frac{1}{{{N_t}}}\tr\left( {\widehat {\bf{R}}_{0m}^0} \right).
\end{align} \end{small}

$\frac{1}{N_t}{\left( {{\bf{h}}_E^l} \right)^H}{{\bf{U}}_{{\rm null},\,l}}{ {{{\bf{U}}_{{\rm null},\,l}^H}}}{\bf{h}}_E^l$ can be expressed as
 \begin{small} \begin{align} \label{eq:hEl_null_app}
& \frac{1}{N_t} {\left( {{\bf{h}}_E^l} \right)^H}{{\bf{U}}_{{\rm null},\,l}}{ {{{\bf{U}}_{{\rm null},\,l}^H}}}{\bf{h}}_E^l \nonumber \\
& \! = \! \frac{1}{N_t}\tr\left( {{\bf{R}}_E^l} \right) \! - \! \sum\limits_{k = 1,k \ne m}^K \! {\tr{{\left( {\widehat {\bf{R}}_{lk}^l} \right)}^{ - 1}}} \frac{1}{N_t} {{{\left( {{\bf{h}}_E^l} \right)}^H}\! \left( {\widehat {\bf{h}}_{lk}^l} \right)\! {{\left( {\widehat {\bf{h}}_{lk}^l} \right)}^H}\! {\bf{h}}_E^l} \nonumber \\
& - \tr{\left( {\widehat {\bf{R}}_{lk}^l} \right)^{ - 1}} \frac{1}{N_t} {{{\left( {{\bf{h}}_E^l} \right)}^H}\left( {\widehat {\bf{h}}_{lm}^l} \right){{\left( {\widehat {\bf{h}}_{lm}^l} \right)}^H}{\bf{h}}_E^l}.
\end{align} \end{small}The asymptotic expressions for $ \frac{1}{N_t}\left({{\bf{h}}_E^l} \right)^H\left( {\widehat {\bf{h}}_{lk}^l} \right){{\left( {\widehat {\bf{h}}_{lk}^l} \right)}^H}{\bf{h}}_E^l$ and $ \frac{1}{N_t}{{{\left( {{\bf{h}}_E^l} \right)}^H}\left( {\widehat {\bf{h}}_{lm}^l} \right){{\left( {\widehat {\bf{h}}_{lm}^l} \right)}^H}{\bf{h}}_E^l}$ in (\ref{eq:hEl_null_app}) are given by
\begin{small} \begin{align} \label{eq:hElwlk}
  \frac{1}{{{N_t}}} {{{{\left( {{\bf{h}}_E^l} \right)}^H}\widehat {\bf{h}}_{lk}^l{{\left( {\widehat {\bf{h}}_{lk}^l} \right)}^H}{\bf{h}}_E^l}} \mathop  \to \limits^{{N_t}\to \infty}   \frac{1}{{{N_t}}} {{\tr\left( {{\bf{R}}_E^l\widehat {\bf{R}}_{lk}^l} \right)}}
 \end{align}  \end{small}and
 \begin{small} \begin{align} \label{eq:hElwlk_2}
  \frac{1}{{{N_t}}}{\left( {{\bf{h}}_E^l} \right)^H}\widehat {\bf{h}}_{lm}^l{\left( {\widehat {\bf{h}}_{lm}^l} \right)^H}{\bf{h}}_E^l \mathop  \to \limits^{{N_t}\to \infty} \frac{1}{{{N_t}}} \Lambda _E^l,
\end{align}  \end{small}respectively.

Substituting (\ref{eq:hE0w0m_app})--(\ref{eq:hElwlk_2}) into (\ref{eq:SINR_E_mul}), we obtain (\ref{eq:e}), which completes the proof.


\begin{thebibliography}{10}
\providecommand{\url}[1]{#1}
\csname url@rmstyle\endcsname
\providecommand{\newblock}{\relax}
\providecommand{\bibinfo}[2]{#2}
\providecommand\BIBentrySTDinterwordspacing{\spaceskip=0pt\relax}
\providecommand\BIBentryALTinterwordstretchfactor{4}
\providecommand\BIBentryALTinterwordspacing{\spaceskip=\fontdimen2\font plus
\BIBentryALTinterwordstretchfactor\fontdimen3\font minus
  \fontdimen4\font\relax}
\providecommand\BIBforeignlanguage[2]{{%
\expandafter\ifx\csname l@#1\endcsname\relax
\typeout{** WARNING: IEEEtran.bst: No hyphenation pattern has been}%
\typeout{** loaded for the language `#1'. Using the pattern for}%
\typeout{** the default language instead.}%
\else
\language=\csname l@#1\endcsname
\fi
#2}}






\bibitem{Wyner1975BST}
{A. D. Wyner}, ``The wiretap channel,'' \emph{Bell Syst. Tech. J.}, vol.~54, pp. 1355--1387, Oct. 1975.


\bibitem{Khisti2010TIT_2} {A. Khisti and G. W. Wornell}, ``Secure transmission with multiple antennas--{Part II: The MIMOME} wiretap channel,''
\emph{IEEE Trans. Inf. Theory},
vol.~56, pp. 5515--5532, Nov. 2010.

\bibitem{Oggier2011TIT} {F. Oggier and B. Hassibi}, ``The secrecy capacity of the {MIMO} wiretap channel,'' \emph{IEEE Trans. Inf. Theory}, vol.~57, pp. 4961--4972, Aug. 2011.



\bibitem{Wu2012TVT}
{Y. Wu, C. Xiao, Z. Ding, X. Gao, and S. Jin}, ``Linear precoding for finite alphabet signaling over MIMOME wiretap channels,''
\emph{IEEE Trans. Veh. Technol.},  vol.~61,  pp. 2599--2612, Jul. 2012.


 \bibitem{Goel2008TWC} {S. Goel and R. Negi}, ``Guaranteeing secrecy using artificial noise,'' \emph{IEEE Trans. Wireless Commun.}, vol.~6,  pp. 2180--2189, Jun. 2008.



 \bibitem{Zhou2012TWC} {X. Zhou, B. Maham, and A. Hj${\o}$rungnes}, ``Pilot contamination for active eavesdropping,'' \emph{IEEE Trans. Wireless Commun.}, vol.~11, pp. 903--907, Mar. 2012.

 \bibitem{Marzetta2008TWC} {T. L. Marzetta}, ``Noncooperative cellular wireless with unlimited numbers of base station antennas,'' \emph{IEEE Trans. Wireless Commun.}, vol.~9, pp. 3590--3600, Nov. 2010.

 \bibitem{Jose2011TWC} {J. Jose, A. Ashikhmin, T. L. Marzetta, and S. Vishwanath}, ``Pilot contamination and precoding in multi-cell TDD systems,''
  \emph{IEEE Trans. Wireless Commun.}, vol.~10, pp. 2640--2651, Aug. 2011.

\bibitem{Yin2013JSAC} {H. Yin, D. Gesbert, M. Filippou, and Y. Liu}, ``A coordinated approach to channel estimation in large-scale multiple-antenna systems,''
\emph{IEEE J. Sel. Areas Commun.}, vol.~31, pp. 264--273, Feb. 2013.

\bibitem{Adhikary2013TIT} {A. Adhikary, J. Nam, J.-Y. Ahn, and G. Caire}, ``Joint spatial division and multiplexing--{The} large-scale array regime,''
 \emph{IEEE Trans. Inf. Theory}, vol.~59, pp. 6441--6463, Oct. 2013.

\bibitem{Wu2014} {Y. Wu, C.-K. Wen, C. Xiao, X. Gao, and R. Schober}, ``Linear precoding for the MIMO multiple access channel
with finite alphabet inputs and statistical CSI,'' to appear in \emph{IEEE Trans. Wireless Commun.}.


\bibitem{Chen2014_2} {X. Chen, J. Chen, and T. Liu}, ``Secure wireless information and power transfer in
large-scale MIMO relaying systems with imperfect CSI,'' [Online]. Available: http://arxiv.org/abs/1407.5355v1.pdf, Jul. 2014.



\bibitem{Zhu2014} {J. Zhu, R. Schober, and V. K. Bhargava}, ``Secure transmission in multi-cell massive {MIMO} systems,''
\emph{IEEE Trans. Wireless Commun.}, vol. 13, pp. 4766--4781, Sep. 2014.



\bibitem{Im2013}{S. Im, H. Jeon, J. Choi, and J. Ha}, ``Secret key agreement under an active attack in MU-TDD systems with large antenna arrays,"
in \emph{Proc. IEEE
  Global. Telecommun. Conf. (GLOBECOM 2013)}, Atlanta, USA, Dec. 2013, pp. 1849--1855.

\bibitem{Wen2013TIT}{C.-K. Wen, G. Pan, K.-K. Wong, M. Guo, and J.-C. Chen}, ``A deterministic equivalent for the analysis
of non-Gaussian correlated MIMO multiple access channels,"  \emph{IEEE Trans. Inf. Theory}, vol. 59, pp. 329--352, Jan. 2013.

 \bibitem{Kailath2000} T.~Kailath, A.~H. Sayed, and B.~Hassibi, \emph{Linear Estimation}. New Jersey: Prentice Hall, 2000.


 \bibitem{Evans2000TIT} {J. Evans and D. N. C. Tse}, ``Large system performance of linear multiuser receivers in multipath fading channels,''
  \emph{IEEE Trans. Inf. Theory}, vol.~46, pp. 2059--2078, Sep. 2000.

 \bibitem{Cho2010} Y.~S. Cho, J.~Kim, W.~Y. Yang, and C.~G. Kang, \emph{MIMO-OFDM Wireless Communications with MATLAB}.
  Singapore: John Wiley $\&$ Sons (Asia) Pte Ltd, 2010.


  \end{thebibliography}
\end{document}